\PassOptionsToPackage{table}{xcolor} 

\documentclass[sigconf]{acmart}

\AtBeginDocument{%
  \providecommand\BibTeX{{%
    \normalfont B\kern-0.5em{\scshape i\kern-0.25em b}\kern-0.8em\TeX}}}

\usepackage{tikz}
\usepackage{multirow}
\usepackage{comment}
\usepackage{float}
\usepackage{environ}
\usepackage{xparse}
\usepackage{hyperref}
\usepackage{wrapfig} 

\usepackage{graphicx}

\usepackage{enumitem}

\usepackage{dirtytalk} 

\renewenvironment{quote}
  {\list{}{\rightmargin=0.27in \leftmargin=0.27in}%
   \item\relax}
  {\endlist}

\newenvironment{sayit}[1]
{\textit{\say{#1}}}

\newenvironment{blockquote}[2]
{\begin{quote}\textit{\say{#1}} (#2)\end{quote}}

\copyrightyear{2024} 
\acmYear{2024} 
\setcopyright{rightsretained} 
\acmConference[ASSETS '24]{The 26th International ACM SIGACCESS Conference on Computers and Accessibility}{October 27--30, 2024}{St. John's, NL, Canada}
\acmBooktitle{The 26th International ACM SIGACCESS Conference on Computers and Accessibility (ASSETS '24), October 27--30, 2024, St. John's, NL, Canada}
\acmDOI{10.1145/3663548.3675622}
\acmISBN{979-8-4007-0677-6/24/10}

\begin{document}

\title{Intersecting Liminality: Acquiring a Smartphone as a Blind or Low Vision Older Adult}


\author{Isabela Figueira}
\affiliation{%
  \institution{University of California, Irvine}
  \city{Irvine}
  \state{CA}
  \country{USA}
}\email{i.figueira@uci.edu}

\author{Yoonha Cha}
\affiliation{%
  \institution{University of California, Irvine}
  \city{Irvine}
  \state{CA}
  \country{USA}
}\email{yoonha.cha@uci.edu}

\author{Stacy M. Branham}
\affiliation{%
  \institution{University of California, Irvine}
  \city{Irvine}
  \state{CA}
  \country{USA}
}\email{sbranham@uci.edu}

\renewcommand{\shortauthors}{Isabela Figueira, et al.}

\begin{abstract}
Older adults are increasingly acquiring smartphones. But acquiring smartphones can be difficult, and little is known about the particular challenges of older adults who are additionally blind or losing their vision. We shed light on the social and technical aspects of acquiring smartphones with vision loss, based on deep qualitative interviews with 22 blind or low vision (BLV) older adults aged 60 and over. Through our grounded theory analysis, we found that BLV older adults experience liminality as they acquire smartphones and transition through re-acquiring smartphones as they become blind, and they can transition through liminality by participating in mutual aid within the blind community. We contribute the notion of ``Intersecting Liminality,'' which explains the marginalizing experience of simultaneously transitioning through vision loss, aging, and technology acquisition. We contend that Intersecting Liminality can serve as a framework that centers the dynamic nature of disability to help our community generate a more nuanced understanding of technology acquisition and more effective assistive interventions.
\end{abstract}





\begin{teaserfigure}
\begin{center}
  \includegraphics[width=\textwidth]{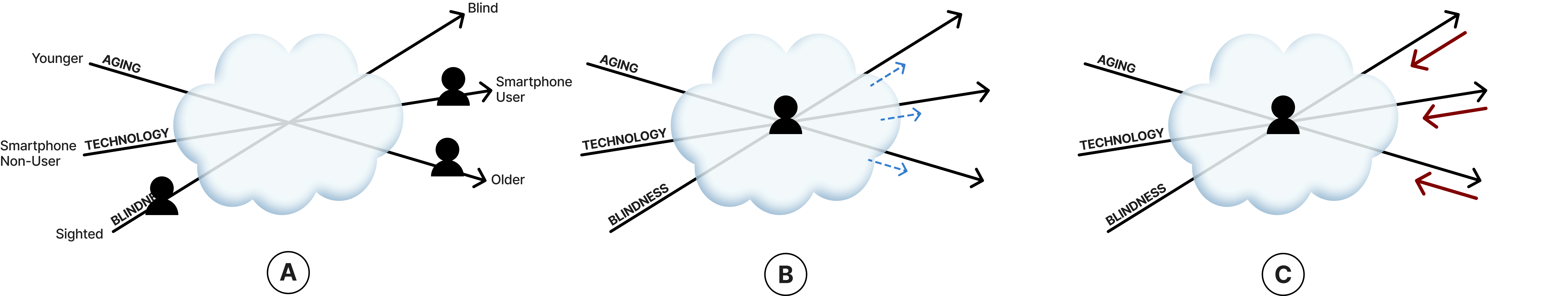}
  \caption{(A) An older adult who has acquired a smartphone 
  (B) gets pushed into liminal space as they acquire blindness and struggle to transition out (blue striped arrows pointing away from clouds. E.g., attending classes, accepting blindness). 
  (C) The Intersecting Liminalities of aging, technology acquisition, and blindness make it difficult to escape (red arrows pointing towards clouds. E.g., lack of Android classes at local centers, constantly learning, daunting learning curve).}
  \Description{Three images describing the story of a person going through the life transition of acquiring blindness while acquiring a smartphone and aging. In each image, in the middle there is a cloud signifying a liminal space. Three axes go through the cloud. Each axis represents a transition the BLV older adult goes through. First axis is aging, from being younger to older. Second axis is technology, from being a smartphone non-user to user. The last axis is blindness, the transition from being sighted to being blind. 
  In image A, there is a person (icon) on the sighted end of the blindness axis, a person on the smartphone user end of the axis, and a person on the older end of the aging axis.
  In image B, one person icon is in the middle of the cloud, signifying that the person is in the liminal space on all axes. On each axis, a blue striped arrow points out and a red arrow points into the liminal space cloud. Blue striped arrows point from the BLV older adult and out of the cloud to signify BLV older adults' efforts to exit the liminal space and factors that support the BLV older adult to get out of the liminal space.
  In image C, one person is in the middle of the cloud. Red arrows point from out of the liminal space towards the BLV older adult to signify factors that push the BLV older adult back into the liminal space.}
  \label{fig:teaser_story}
  \end{center}
\end{teaserfigure}


\maketitle

\section{Introduction}

76\% of the 56 million older adults aged 65 and older in the USA \cite{us_census_bureau_exploring_2023} are using a smartphone \cite{sidoti_mobile_2024} in 2024. Acquiring a smartphone in older age is important for accessing online information \cite{sidoti_mobile_2024}, supporting aging in place \cite{caldeira_i_2022}, or increasing communication with healthcare professionals \cite{ciano_importance_2020}. 
We know that mobile technology acquisition can be challenging for older adults due to interface complexity \cite{mcgaughey_motivations_2013} and a lack of instruction manuals \cite{leung_how_2012}, for example. 
In response, interventions have been proposed to support older adults in learning smartphones \cite{wilson_help_2018} and locating features on mobile interfaces \cite{yu_supporting_2020}. 
However, what about those who additionally experience disability, specifically vision loss, in older age? 

In 2017, in the USA, of those over the age of 65, approximately 4.2 million people were low vision, and another 800,000 were blind\footnote{In this paper, we use the term low vision to indicate visual acuity of $\leq20/40$, and blind to indicate visual acuity of $\leq20/200$} \cite{cdc_prevalence_2023, flaxman_prevalence_2021}. 
For blind people, ``mastering these devices is an arduous and long task'' \cite{rodrigues_getting_2015}, since accessing a smartphone entails using features that are often buried in the phone, unfamiliar to the general population, and difficult to configure \cite{franz_perception_2019}. Despite this growing body of work on blind and low vision (BLV) older adults and smartphones~\cite{figueira_smartphone_2023, piper_technology_2017, piper_understanding_2016, fuglerud_social_2021}, there is insufficient research on how older adults who are BLV acquire smartphones.


Acquiring disability and becoming older have been described as life transitions~\cite{salovaara_information_2010, nilsson_feeling_2000, buscherhof_abled_1998}. For example, adults can experience multiple transitions such as changes in health and retirement status as they enter later life~\cite{salovaara_information_2010, grenier_transitions_2012}, and maintaining a positive attitude has been found to support the transition into disability~\cite{senra_psychologic_2015}. 
With life transitions, people can experience liminality, as they do not completely belong to either the pre- or post-transition states. The word liminal is defined as ``relating to a transitional or initial stage of a process'' and ``occupying a position at, or on both sides of, a boundary or threshold'' \cite{stevenson_liminal_2015}. 
There have yet to be reports of the liminal experiences of those who go through transitions of acquiring blindness and smartphones in older age, and how experiencing these multiple transitions simultaneously introduces additional complexities in identity shifts during smartphone acquisition.

To address this gap, we conducted a qualitative study to answer the following research questions:
\begin{enumerate}[label=(RQ\arabic*), leftmargin=1.25cm]
    \item How do older adults who are BLV acquire smartphones?
    \item How do older adults who are transitioning into BLV acquire smartphones?
\end{enumerate}
We report on in-depth semi-structured interviews with 22 participants who were 60 years of age or older, who identified as being blind or low vision, and who used a smartphone. Through a constructivist grounded theory analysis ~\cite{charmaz_constructing_2014}, using theoretical sampling and abductive reasoning, we read our data through a lens of life transition and liminality, which revealed three main themes: 
(1) Older adults who have early life blindness\footnote{In this paper, we define early life BLV to mean blindness or low vision that was acquired before the age of 60. We use early life BLV, early life blindness, and early life vision loss interchangeably throughout the paper.} experience liminality as they acquire smartphones, 
(2) Older adults with later life blindness experience liminality as they transition through re-acquiring smartphones they already owned as they acquire blindness, and 
(3) BLV older adults transition through liminality by engaging in mutual aid with the blind community as they acquire smartphones. 

Based on our findings, we present \textit{Intersecting Liminality}, a framework for articulating the ways in which multiple liminalities–between younger and older, between sightedness and blindness, and between smartphone non-user and user–can intersect and compound (section \ref{sec_51_intersecting_liminality}). 
We argue that in order to holistically understand how older adults with disabilities acquire technology, and to design effective interventions, the multiple liminalities they experience must be considered as inextricably linked.

\section{Related Work}

Scholars have studied older adults and technology for decades from many perspectives, such as life course sociology that focuses on life stages \cite{elder_age_1975, mayer_new_2009, settersten_it_2009, gilleard_connecting_2016}, lifespan psychology that focuses on the individual 
\cite{erikson_childhood_1963, harley_growing_2018, settersten_it_2009, gilleard_connecting_2016}, social gerontology \cite{neves_can_2019}, medicine \cite{cotten_impact_2013, piper_supporting_2013, chen_understanding_2021}, feminist theory \cite{brewer_challenging_2021, lazar_adopting_2021}, and more. 
Narratives about aging describe a growing digital divide \cite{ball_physicaldigital_2019, neves_old_2018, mcgaughey_motivations_2013} and a lack of technological savvy \cite{fuchigami_assistive_2022}. Additionally, older adults increasingly face social isolation \cite{seifert_double_2021, baecker_technology_2014,shinde_designing_2024} 
and digital exclusion \cite{coleman_engaging_2010} as communication moves increasingly online. 
While older adults have traditionally been considered passive users of technology~\cite{brewer_challenging_2021, nimrod_fun_2011, chang_age_2015, Uden-Kraan_self_2008}, recent research has challenged this perspective, considering older adults as actively passive users \cite{knowles_wisdom_2018, ellison_why_2020}, online content creators \cite{brewer_tell_2016, tang_towards_2023}, and caregivers~\cite{brewer_challenging_2021}. At the same time, there has been a shift in discourses around older adulthood from foregrounding deficits~\cite{vines_age-old_2015, mclaughlin_using_2009, worden_making_1997, becker_web_2004}–lack of capability–to foregrounding assets \cite{morrow-howell_more_2023, lee_designing_2023, harrington_examining_2022, vines_age-old_2015}.

In this work, we draw on perspectives from social gerontology, namely the concepts of life transitions and liminality, and we concur that older adults are active and capable in their social and technical lives.

\subsection{Life Transitions, Liminality, and Technology}

Across the life course, a person experiences many life transitions such as becoming an adult, getting married, growing a family, etc. \cite{grenier_transitions_2012, salovaara_information_2010}.
As one goes through a life transition, they go through a liminal period. van Gennep~\cite{van_gennep_rites_1909} proposed that life transitions comprise a three step process beginning with separating from a previous identity or environment, followed by a liminal or transitional period, and ending with being incorporated into a new social context. Turner~\cite{turner_liminality_1969, turner_betwixt_2001} 
further theorized liminal space from an anthropological perspective. As a person goes through a transition, they are separated from social structure and enter a state of ambiguity in which one becomes a ``blank slate'' to prepare for learning new knowledge from their new group as one incorporates \cite{turner_liminality_1969}. He unpacks what it means to experience liminality, and how there is a symbolism attached to the ``liminal persona'' that can take on both bizarre and negative connotations, since one is viewed as other and being outside of known social groups during the transition between them \cite{turner_betwixt_2001}. 

Becoming an older adult and acquiring disability have both been identified as life transitions that carry with them the features of liminality described above. As individuals age, they experience changes in various aspects of life such as health, retirement status, financial status, housing situation, and amount of social interaction~\cite{salovaara_information_2010, grenier_transitions_2012, leibing_liminal_2016, west_enjoying_2017, rudman_why_2016}. These constitute a process of becoming, rather than a state of being older, one in which individuals can resist, move through, or occupy liminal space indefinitely \cite{hockey_social_2003}. In the latter case, those who are moving from the third to fourth age (i.e., the oldest old) can be said to experience ``persistent liminality'' 
as they grow increasingly isolated and housebound with age \cite{nicholson_living_2012}.  Similarly, acquiring disability can be conceptualized as an identity transition from a non-disabled to a disabled person, a process in which a ``new sense-of-self'' comes into being. This can comprise multiple stages including denial and grieving, building a social support system, successfully being reintegrated into the community, and finally accepting disability and returning to a quality life~\cite{buscherhof_abled_1998}. 

There has been increased attention on liminality in social computing research \cite{haimson_life_2019}. Researchers have conceptualized liminality in terms of physical spaces \cite{freire_ravenet_2024, liedgren_liminal_2023, macdonald_designing_2015}, digital spaces \cite{ringland_place_2019, martinez_liminal_2011}, and as being between learning and mastery \cite{eckerdal_limen_2007}. Researchers have also studied how people appropriate technologies during the process of life transitions \cite{kluber_designing_2020, wolf_i_2022} and how online communities such as social media support people experiencing liminality during life transitions \cite{semaan_transition_2016, morioka_identity_2016, haimson_social_2018}. For example, Haimson~\cite{haimson_social_2018} detailed how multiple social media sites work as ``social transition machinery'' to support people experiencing life transitions \cite{haimson_social_2018}. Semaan et al.~\cite{semaan_transition_2016} found that social media helped veterans experiencing multiple life transitions develop “identity awareness” as they received support from other veterans to reintegrate into civil society \cite{semaan_transition_2016}.
Researchers have also called for designing technology for people who are socially isolated and thus in a perpetual liminal space, since they are in between or outside of communities and not integrated into social practices and patterns \cite{jensen_digital_2020}.

We adopt Turner's process approach to liminality that centers becoming and being othered.
Liminality has not yet been conceptualized or applied in accessible computing in the field of HCI, as virtually no work has documented the experiences of disabled people experiencing life transitions. 

\subsection{Older adults' smartphone acquisition and adoption}

Scholars have studied older adults' technology adoption from multiple angles, such as adopting social media \cite{brewer_challenging_2021, 
cotten_social_2022, bell_examining_2013, hope_understanding_2014}, online shopping \cite{seo_power_2023},  online banking \cite{jin_i_2022, jiang_bringing_2022, 
latulipe_unofficial_2022}, 
communication technologies~\cite{neves_my_2015}, wearables for health management \cite{pang_technology_2021}, voice assistants \cite{upadhyay_studying_2023}, and smart mobile devices \cite{piper_understanding_2016}. 
Several studies mentioned that social support is crucial for adopting computing technology, but also presents challenges, such as living far from family \cite{seo_power_2023, piper_understanding_2016, tang_i_2022, neves_coming_2013}.  
Several factors can affect older adults' adoption of technology in addition to age \cite{hanson_influencing_2010, knowles_older_2018, schroeder_older_2023, beneito-montagut_emerging_2022}, including identity \cite{smith_user_2019}, confidence \cite{an_understanding_2022}, and stigma associated with disability or aging \cite{astell_thats_2020}.

Prior work regarding older adults and smartphone adoption has focused on motivations to adopt a smartphone (e.g., the desire for independence) \cite{mcgaughey_motivations_2013, conci_useful_2009} and learning preferences (e.g., the desire for instruction manuals) \cite{leung_how_2012, piper_understanding_2016, sharifi_senior_2023}. One popular framework for explaining how technology is adopted is the technology acceptance model (TAM) ~\cite{davis_perceived_1989}, which has been extended for older adults ~\cite{renaud_predicting_2008, chen_gerontechnology_2014}. However, these frameworks focus on usability, leading researchers to question whether they are too rigid to apply to older adults \cite{neves_my_2015, franz_time_2015}. For example, in a study of older adults adopting a tablet-based communication technology, Neves at al. conclude that \say{TAM is limited as a theoretical model since it neglects the interplay of social context, human agency (individual choices), and inherent properties of technology} \cite{neves_my_2015}.

While a great deal of work focuses on how BLV people use or access mobile devices (e.g., \cite{barbareschi_social_2020,jain_smartphone_2021,azenkot_exploring_2013,szpiro_how_2016,kane_freedom_2009,tigwell_emoji_2020}), 
few papers have focused on BLV people's adoption of smartphones in our research community. 
In 2015, Rodrigues et al.~\cite{rodrigues_getting_2015} studied how five novice blind people adopt, learn, and use an Android smartphone over an 8-week period. Following the adoption period, they found that although learning smartphones was a long and difficult process, participants were resilient and willing to keep learning, since they valued the benefits the smartphone could provide. 
Very little research has highlighted the combination of aging, blindness, and smartphone acquisition, with three notable exceptions. 
Piper et al.~\cite{piper_understanding_2016} studied older adults over the age of 80, some with vision loss, and found that initial setup and learning of smart mobile devices, mostly tablets, was difficult due to lack of access to one-on-one support. In another study, Piper et al.~\cite{piper_technology_2017} studied how older adults with later life vision loss learn and use computing technology to communicate, where two participants were smartphone users. 
Only one work focused specifically on smartphone acquisition challenges of BLV older adults–on a small scale \cite{figueira_smartphone_2023}. 
Our study adds to this corpus by being the first to document the unique experiences of older adults who are blind or losing their vision as they acquire a smartphone.

\section{Methods}

\begin{table*}[ht]
\small
\begin{tabular}{| p{0.03\linewidth} 
               | p{0.03\linewidth} 
               | p{0.4\linewidth} 
               | p{0.25\linewidth} 
               | p{0.08\linewidth} 
               | p{0.06\linewidth}|}
              \hline

\textbf{ID} &
  \textbf{Age} &
  \textbf{Vision Ability} &
  \textbf{Current Smartphone Model(s)} &
  \textbf{Smartphone Use (Years)} &
  \textbf{Gender} \\ \hline
E1  & 64                       & Low Vision                                          & iPhone 12 Pro Max                       & 6 to 10  & Woman \\ \hline
E2  & 72                       & Low Vision                                          & iPhone 14 Pro, BlindShell Classic 2     & 11 to 15 & Woman \\ \hline
E3  & 70                       & Blind                                               & iPhone SE                               & 1 to 2   & Woman \\ \hline
E4  & 60                       & Low Vision (light perception)                       & iPhone SE                               & 6 to 10  & Woman \\ \hline
E5  & 70s & Totally Blind                                       & iPhone SE 2022 third generation                & 5 to 10  & Woman \\ \hline
E6  & 68                       & Totally Blind                                       & iPhone 12                               & 6 to 10  & Woman \\ \hline
E7  & 69                       & Totally Blind                                       & iPhone 12 Mini                          & 11 to 15 & Man   \\ \hline
E8  & 76                       & Totally Blind                                       & iPhone with home button                 & 1 to 2   & Man   \\ \hline
E9  & 80                       & Light Perception                                    & two iPhone 6S, BlindShell Classic 2     & 5 to 10  & Man   \\ \hline
E10 & 65                       & Totally Blind                                       & iPhone 6s                               & 11 to 15 & Man   \\ \hline
E11 &
  75 &
  very Low Vision, uses Braille \& white cane &
  iPhone with home button, 3 years old &
  3 to 5 &
  Man \\ \hline
E12 & 62                       & Low Vision. uses VoiceOver on phone                 & iPhone 14 Pro Max                       & 15+      & Man   \\ \hline
E13 & 68                       & Totally Blind                                       & iPhone 14 Pro Max                       & 11 to 15 & Woman \\ \hline
E14 & 83                       & Totally Blind                                       & (Android) Pixel 5A                      & 5 to 10  & Man   \\ \hline
E15 &
  65 &
  Totally Blind &
  (Android) Google Pixel 7 Pro, Samsung Galaxy A53 5G &
  5 to 10 &
  Man \\ \hline
E16 & 61                       & Low Vision / legally blind                          & (Android) Google Pixel 3 XL                       & 10 to 15 & Man   \\ \hline
L1  & 83                       & Totally Blind                                       & iPhone 6, SE, BlindShell Classic 1 \& 2 & unknown  & Man  \\ \hline
L2 &
  73 &
  almost Totally Blind, can see blurry shapes and movement &
  Android 12, Motorola G Power &
  15+ &
  Man \\ \hline
L3  & 66                       & Visually Impaired, light perception                 & iPhone 12                               & 15+      & Man   \\ \hline
L4 &
  72 &
  almost Totally Blind, some peripheral vision \& light perception &
  iPhone 14 Pro &
  5 to 10 &
  Man \\ \hline
L5  & 76                       & Low Vision, advanced glaucoma                       & iPhone 14 Plus                          & 5 to 10  & Man   \\ \hline
L6  & 75                       & Low Vision. depending on light can see some details & iPhone 13                               & 11 to 15 & Man   \\

\hline
\end{tabular}%
\caption{Demographics. Information is self-reported by participants. IDs that begin with E represent participants with early-life vision loss before the age of 60, and IDs beginning with L represent participants with later-life vision loss at age 60 or older.}
\label{tab:demographics}
\end{table*}

\subsection{Participants}

We recruited 22 participants who identified as being blind or having low vision, who use a smartphone, and who are aged 60 or over. Participants were recruited from local independent living centers, online forums about technology, and snowball sampling.

To maintain confidentiality of our participants' information, we replaced their names with IDs. Participants' ages ranged from 60 to 83 years. Seven participants were women, and 15 were men. Out of 22 participants, six identified as having lost vision later in life at age 60 or later (IDs begin with L), and 16 had been blind since earlier in life (IDs begin with E). Of the six participants with later life vision loss, all used smartphones in some capacity prior to losing vision, although two described re-learning the smartphone from scratch after vision loss. Most participants (n = 18) used iPhone. Of these 18, three participants used a combination of iPhone and BlindShell\footnote{BlindShell is a smartphone designed for blind people. \url{https://www.blindshell.com}}. Four participants used Android. Most participants (n = 18) had been using their smartphones for at least five years and had used several models of smartphones in the past. Refer to \autoref{tab:demographics} for details.

\subsection{Procedure}

We conducted audio-recorded, semi-structured interviews with 22 participants over a phone call or Zoom, according to their preferences. Interviews were conducted over a period of 10 months in 2023. Each interview lasted between 50 and 96 minutes (average of 76 minutes, total of over 26 hours). Prior to the interview, we emailed participants a study information sheet for review and acquired verbal consent at the beginning of the interview. Participants also filled out a pre-interview demographic survey. Participants were compensated at a rate of \$40 per hour in the form of a gift card. The study was approved by the researchers' Institutional Review Board (IRB). 

Interview questions covered topics such as smartphone opinions, accessibility challenges experienced, how the smartphone was used for social communication, and  kinds of resources they utilized when learning smartphones. After the first five interviews, the first author engaged in theoretical sampling \cite{charmaz_constructing_2014} by adding more questions about aging  and social communication on the smartphones and by asking targeted questions about losing vision with a smartphone to participants with later life vision loss. Later in the study, the researchers also specifically recruited more Android users.

\subsection{Analysis}

All interviews were transcribed by either the researchers or professional transcribers. The transcripts  were checked  for quality and accuracy by the researchers. We followed a constructivist grounded theory method \cite{charmaz_constructing_2014}, starting with open coding on each of the transcripts. The first two authors open coded data in parallel and met frequently to discuss emerging themes, such as the importance of social support, social exclusion on the smartphone, differences in expectations vs. reality regarding the smartphone, etc. They engaged in focused coding together, frequently meeting to discuss ideas emerging across participant data. As the authors compared focused codes from participants with later life and early life blindness, we recognized the concept of identity transition and revisited the literature to review the concepts of life transition and liminality. We decided to take an abductive approach to the remaining analysis, in such a way that these extant theoretical concepts \say{earned their way} into our final narrative \cite{charmaz_constructing_2014}. All three authors engaged in theoretical sorting through diagramming \cite{charmaz_constructing_2014}, and in particular, affinity diagramming~\cite{lucero_using_2015} using FigJam~\cite{figjam}, to streamline the core ideas into axial codes. The headings and subheadings of our Findings map to the axial and focused codes of our analysis (see \autoref{tab:codes}).

\begin{table}[htbp]
\resizebox{\columnwidth}{!}{%
\begin{tabular}{|p{4.4cm}|p{5.25cm}|}
\hline
\textbf{Axial Codes} & \textbf{Focused Codes} \\ 
\hline
\multirow{2}{4.4cm}{Liminality of Acquiring a Smartphone as a Blind Older Adult} & Becoming a Smartphone User \\
 & Social Exclusion \\ 
\hline
\multirow{3}{4.4cm}{Liminality of Re-Acquiring a Smartphone While Acquiring Blindness as an Older Adult} & Becoming Blind \\ 
 & Burden of Re-Acquiring the Smartphone \\ & \\
\hline
\multirow{3}{4.4cm}{Transitioning via Mutual Aid} & Going for the Phone, Gaining Community \\
 & Cycle of (non)Acceptance \\ 
 & The Hole of Later Life Vision Loss \\ 
\hline
\end{tabular}%
}
\caption{Findings Structure}
\label{tab:codes}
\end{table}
\section{Findings}

We found that BLV older adults experience liminality as they acquire a smartphone, experiencing social exclusion as a result (section \ref{sec_41}). Older adults who become blind later in life experience multiple liminalities as they both acquire blindness and re-acquire smartphones they used when sighted (section \ref{sec_42}). Finally, BLV older adults transition through liminality via mutual aid with the blind community as they (re)acquire smartphones (section \ref{sec_43}). We see especially in sections \ref{sec_42} and \ref{sec_43} that BLV older adults experience compounding marginalization, rooted in the multiple liminalities of acquiring blindness, acquiring a smartphone, and aging; from this, we propose Intersecting Liminality, a framework we elaborate in the Discussion (section \ref{sec_51_intersecting_liminality}).

\subsection{Liminality of Acquiring a Smartphone as a Blind Older Adult} \label{sec_41}

\subsubsection{Becoming a Smartphone User} \label{sec:41acquiring_phone}
As BLV older adults switch from a feature phone to a smartphone, they may experience conflicting understandings of their own identities as smartphone users, finding themselves "stuck" in transition. 

\paragraph{From Unimaginable to Indispensable}
Many could not imagine themselves as smartphone users, especially since stories about smartphones didn't align with their needs. Multiple participants originally questioned how they could navigate a touchscreen without the tactile buttons characteristic of their feature phones. Additionally, E3 recounted how \sayit{you're only told \say{you need to get an iPhone}} (E3), but she did not understand clearly why a blind person should get an iPhone or how it could support her as a blind person: \sayit{Never the why... If they do a why, it's just you can't relate to it because it's like, \say{well, how does that benefit me as a blind person?}} (E3). But, once they acquired a smartphone, they perceived their smartphones as indispensable in their daily lives. Multiple participants (e.g., E4 and E12) described the smartphone as \sayit{something that, once you did it, you kinda wondered, `How did I ever live without this?'} (E4) and even feeling \sayit{lost} (E8) without their smartphones.
E1 was reluctant to get a smartphone since she \sayit{never wanted to be the type of person who runs around with their phone all the time. Always on the phone talking} (E1). However, \sayit{now I cannot go anywhere without my phone, traveling to the doctor or anywhere because I need my phone} (E1). 
Once BLV older adults acquired a smartphone, their identities shifted from skepticism to not being able to imagine life without one. The transition to smartphone user is not as simple as it seems. Participants worked hard to learn their smartphones, achieving different levels of mastery. While attempting to learn,  participants are repeatedly dragged back into a state of ambiguity.

\paragraph{Perpetual State of Learning}
Our participants described that acquiring a smartphone posed a big learning curve, after which they were trapped in a state of endless learning, due to accessibility challenges and the smartphone's vast depth–always the beginner, never the expert.

Many participants described hearing they should get an iPhone, but these stories neglected to mention both an inaccessible setup process, often requiring someone sighted \sayit{to make it useful for a blind person} (E9) and a \sayit{difficult} (E1) learning curve. For instance, E3 described tripping up on specialized terminology: \sayit{sometimes I don’t even know the terminology. Like it will say something about... the rotor, do something with the rotor. I'm like, `What the hell is the rotor?!' (laughs) I don't know what that is!} (E3). After the large learning curve, participants described smartphones as still having depth \sayit{like an ocean} (E8), resulting in \sayit{unlimited} (E8) learning. E3 described that \sayit{there’s just so much that I don't know that I should know. ... [although] I've had this phone a little over a year} (E3). 
Regardless, E2 still encouraged others to get a smartphone, due to its ability to support being blind–just \sayit{don't try to learn everything on it} (E2). For some BLV older adults, learning a smartphone felt insurmountable due to gesture challenges. E9 described not feeling successful at using a touch screen phone and thus switched to BlindShell: \sayit{I think there’s two different types [of smartphones]. And I am unable to use one of them with any great degree of success. And that is the type that is a flat screen type} (E9).

Thus, participants felt like they were neither a novice nor an expert, and were stuck learning, relearning, and re-acquiring the same smartphone, where it's difficult to reach a state of mastery. E8 described feeling both like an expert and like he was still learning everything on the smartphone: \sayit{Well, kind of both. I feel those things I've been using I think I learned but I have a lot to learn} (E8). Similarly, E6 described having \sayit{been at it for, ... let's see, five, six years. ... I feel like I'm pretty good with it. Yet I still have issues with iPhone. But I am learning} (E6).

\paragraph{Back to Square One}
Even if participants felt proficient at using their smartphones, thus being past the transition phase to smartphone user, unforeseen external factors such as unreliable smartphone accessibility pushed them into a liminal state.
Sometimes, participants questioned their ability to use their smartphones due to issues of unknown origins. When issues or \sayit{glitches} (E4) occurred, such as iOS and app updates (as described by nearly all participants), some described themselves as \sayit{not smart enough to figure out how to do a workaround} (E3). Additionally, since \sayit{sometimes you don't know if it's a bug} (E13), participants may fault their abilities for causing the issue:
\begin{quote}
    \sayit{Is it by design, sort of broken by design? Well, we don't know yet, because your ability to use it is in question, because you just found it. ... So, I wait until I'm sure that whatever I’m doing isn't the problem.} (E13)
\end{quote}
As a result, some users were compelled \sayit{to, out of necessity, become a more of a technology geek than your natural inclinations} (E7), to deal with smartphone issues.

\subsubsection{Social Exclusion} \label{sec412}
Most participants described largely using their phones for social connection and communication. However, having to conform to less accessible communication preferences of sighted people (i.e., text messaging rather than phone calls) to stay connected with friends and family amplified social isolation and diminished typical support systems, pulling BLV older adults back into liminality as smartphone users. 

\paragraph{``Just Call Me'' and ``Just Let It Go''}
Participants described being socially excluded from conversations through breakdowns in digital communication (e.g., text messaging) when conversing with younger, sighted family and friends.
Voice calls were the preferred mode of communication for our participants, who were familiar with this functionality from feature phones and landlines. Yet, staying socially connected often meant participating in group SMS chats that posed accessibility barriers. Since emojis and images \sayit{litter} (E1) group chats, text conversations tended to sound meaningless and be difficult to follow. For example, during holidays, when group chats exploded with text messages containing emojis such as \sayit{firecracker, skyrocket, [gibberish]} (E10), participants described preferring to be left out of group chats, since they couldn't fully participate anyway: \sayit{My phone is just going on, and on, ... for nothing. ... thankfully, they've kind of dropped me out of that thing, so I don't have to through the holidays, or family parties, or whatever. ... Just call me} (E10).

As close family tended to better understand and accommodate disability in smartphone communication, accessibility issues  most commonly arose out of the disability awareness gap with more casual friends and the generational gap with younger relatives. E1 explained that \sayit{people don't understand your level of vision impairment. (sighs) They send it anyway.} E6 described having to text, especially with younger relatives: \sayit{When I have a nephew or a niece having a birthday or whatever I used to just text them because... they don’t answer the phone. They'd rather do text} (E6). Her close relatives understood her access needs, but others did not, causing her to be excluded: \sayit{My yoga class and my other relatives, nephews and nieces, they all communicate back and forth. ... they just do their own thing. Ok fine, I just delete them} (E6). 

Like E10 and E6, many others described having to accept that they could not participate in certain conversations. Disheartened, they described reluctantly having to \sayit{just move on} (E1) and \sayit{just let it go} (E1). They ultimately felt devalued, powerless, and resigned to being left behind: \sayit{It took me a while, but I just accept it. ... I'm glad that they communicate with one another. I'm just the auntie; what am I? I'm not a cousin, like their cousins. No big deal} (E6). Similarly, E11 \sayit{put[s] it aside. ... I've got more control over my life than getting upset over something that I don't have control over.} 

Although BLV older adults described wanting to participate in group conversations via text, many felt forced to not participate and ultimately were excluded.

\paragraph{Speaking Different Smartphone Languages}

On top of the social exclusion of socializing on the phone with friends and loved ones, BLV older adults experienced a communication barrier when talking about the smartphone with sighted users. This precluded them from giving assistance to or getting help from typical support systems. E7 shared that he and his wife \sayit{always have this communication barrier} (E7) when he wants to help her with her smartphone since his smartphone interface with a screen reader is different:
\begin{quote}
    \sayit{She’ll say `how do you do whatever.' And I'll say `I can't explain it to you, because you're not using VoiceOver, and I don't understand how to do it.' ... Because VoiceOver verbalizes everything. But when there's a button that says `delete,' I always say, `look to the delete button.' But, visually there isn't a delete button, it’s a trash can.} (E7)
\end{quote}
Additionally, when E3 wanted to get assistance from a sighted person on her phone, she described having \sayit{to take the VoiceOver off. ... That doesn't help me. (laughs) But that's the only way they can help me because they don't know VoiceOver} (E3). Thus, she was barred from following along with the sighted helper. The way around this communication barrier is learning both the sighted and blind version of the smartphone, which participants described having to do, \sayit{out of necessity, or we'll just be that blind person in the corner who everybody feels sorry for. Which I am not interested in being} (E7). Thus, BLV older adults turn to the blind community for support (section \ref{sec_43}).

\subsection{Liminality of Re-Acquiring a Smartphone While Acquiring Blindness as an Older Adult} \label{sec_42}

As older adults experience vision loss later in life, they go through additional liminality as their ability and identity shift from sighted to blind. They experience breakdowns in social life and support systems for technology that can lead to them feeling stuck and isolated. Six participants 
experienced later life vision loss.

\subsubsection{Becoming Blind} \label{sec_421}
The transition from being sighted to blind in later life derailed older adults' lives and caused an identity shift or crisis, as they experienced changes in their vision, social circles, support systems, and daily lives in general. Accepting blindness and embracing the blind community was one way of moving forward in life, while not accepting it could result in being stuck in more precarious liminal space between sighted and blind.

\paragraph{Being ``Reborn'' into Isolation}

Some participants described having to relearn everything in life again, including how to walk, balance, navigate in the world, and interact with people as they acquire a new disabled identity. For instance, L6 equated sight loss to being \sayit{reborn} because he \sayit{had to learn how to walk again. ... to now feel my way around with a cane, so I have to walk slower, and carefully. ... to learn to use my ears more} (L6). Participants expected that they would be done learning at their age: \sayit{I didn't think that at my age, in my fifties and sixties, ... oh gosh it would be nice to ... stop working and kind of relax. But it's like, now it's back to grade school} (L2). However, as they acquired blindness, they described having to \sayit{put a lot of effort into any of it. Any of the accessibility things in order to recover your life} (L2). On top of relearning physical navigation, L3 described re-entering society with a new disabled identity: \sayit{People do treat you differently. ... It was different to see the world through the perspective of a disabled person. Because until you've been there you can't conceptualize it} (L3).

Older adults experiencing vision loss become isolated from their social circle. L3 described having a social group that is \sayit{pretty small. But I talk to the people at the [local independent living center] and other classmates} (L3). He shared that \sayit{not being able to get in a car} (L3) reduced in-person visits with family. L1 also described losing contact with friends due to their passing or self isolating due to feeling \sayit{sort of embarrassed with the situation that I'm in} (L1). L1 and L3 described being \sayit{homebound} (L3) due to their vision loss, leading to further isolation. While L3 goes on weekly walks, L1 shared, \sayit{I really don't go anywhere. I usually stay in my apartment, and people come here} (L1). 

Support systems that previously worked for smartphone learning no longer worked after vision loss. As participants continued acquiring blindness, their family's ability to support them declined, since they didn't understand assistive technology: 
\sayit{my son and my daughter are a lot more tech savvy than myself, but ... not in these accessibility things. They don't have any specialty in accessibility, or special knowledge there. ... I don't always have somebody} (L5). 

\paragraph{Accepting Blindness and the Blind Community} \label{sec:accept_blindness}

Participants described needing to accept blindness in order to move forward in life. L3 characterized vision loss as a process of acceptance: \sayit{Losing your sight is ... sort of like the seven steps. Where first there's denial and then acceptance, and so it took a while, just to get used to} (L3). For L6, accepting blindness was an active endeavor and was the key to re-joining society and continuing to live:
\begin{quote}
\sayit{You can't give up, though. ... I don't have a choice. It's either this, or the graveyard, and I don't wanna die yet; I'm not ready. I'm not going to sit at home and become a couch potato, so, I'm going to get out and walk around and join the rest of the world.} (L6)
\end{quote}
L6, like other participants, perceived accepting blindness and, in particular, accepting the blind community, as critical to overcoming social isolation that can stem from vision loss:
\begin{quote}
\sayit{Being blind, it can be a lonely life, if you let it be. And by having the phone, it allows you to be able to open up your life to other people, and let other people into your life. ... You've got to get to know your [blind] community. Can't be a hermit.} (L6)
\end{quote}
We see here and in the sections that follow that the supports provided by the blind community to successfully adopt a smartphone further connect the blind older adult to \sayit{the rest of the world} (L6). Moreover, adoption of the accessibility features of a smartphone \textit{is} a form of acceptance of blindness.


This is precisely why L1 was unwilling to adopt the smartphone or other assistive technologies. L1 described that learning these devices would feel like giving up hope of regaining vision: 
\begin{quote}
\sayit{If I try to have somebody come in and teach me how to use a cane, it's like I'm giving up that goal of mine, which is to get my vision back, ... so I'm not pushing the other resources available like taking classes on how to use a computer.} (L1)
\end{quote}
Further, he described a compounding sense of hopelessness that, even if he did accept his blind identity, he might never be able to achieve smartphone adoption:
\begin{quote}
\sayit{Even if I accepted the fact that I'll be blind forever, it's very time consuming to go to these places [blind organizations] and learn things ... so if I try to really spend my time really trying to learn all these different things, its like I'm giving up my hope.} (L1)
\end{quote}
L1 is resigned to residing in liminal space,  as a man who is neither sighted nor blind and who is no longer able to use his adopted smartphone. The finality of "blindness" and the foreign complexity of assistive technology make transition seem impossible. If L6 is right, and connecting with the blind community to learn the smartphone is the way out of isolation, then L1 has lost much. 
\subsubsection{Burden of Re-Acquiring the Smartphone} \label{sec_422}
Our participants reported having to relearn how to use their existing smartphone or switch to another phone altogether. We refer to this as "re-acquiring" the smartphone, to emphasize the additional labor and the push away from technological confidence and back into a liminal state of learning experienced by older adults who are acquiring blindness. 


\paragraph {Learning to Walk Again}
While L5 chose to make a switch to Apple's iPhone when he acquired vision loss, and did so with relative ease with resources provided by local organizations, this was not the case for the other five participants who lost vision later in life. Four participants (L2, L3, L4, L6) each kept the same smartphone they had previously. Although they had in depth knowledge of the smartphone system as a sighted person, these four participants each needed to relearn the smartphone after acquiring blindness, thus re-acquiring their smartphone. 

Relearning the smartphone posed a \sayit{daunting} (L2) task, one akin to learning \sayit{how to walk all over again} (L6) or even \sayit{a language that is different with my phone, using VoiceOver and finger gestures} (L6). 
Further, since L3 was \sayit{just basically} using his \sayit{iPhone for emails, text and some web surfing} (L3) when he acquired blindness, he described a double labor of relearning the smartphone and learning accessibility functionality:
\blockquote{There was a very steep learning curve, not only in terms of the accessibility functionalities. But the iPhone in general. My wife is already telling me, ‘you use about two percent of the capacity of that thing.’ ... so I had to learn both the regular stuff about iPhones. And the accessibility. ... I’m still learning things as we speak. I had a class today and it’s going to continue for a while.}{L3}

Aware of the massive effort ahead, some participants described proactively preparing to use their smartphone once their vision loss progressed. While L4 \sayit{had to learn things the hard way} since his \sayit{blindness occurred very suddenly} (L4), L2 could prepare his smartphone for accessibility since he acquired blindness gradually: \sayit{I actually set up my voice assistant before I lost my sight completely when I knew things were going that way} (L2). L3 had low vision, but he shared that becoming blind was \sayit{a possibility for me. So I want to learn [VoiceOver] as much as possible so should that day ever come, I’m not starting from square one} (L3).

For these participants, challenges of age and vision loss intersected, resulting in a feeling of having to learn forever or racing against time to regain technical fluency. Participants noted that it can take older adults \sayit{longer to learn things like [technology]} (L6), and vision loss means you \sayit{have to memorize everything. ... you have to repeat it over and over and over again} (L2). 
Paired with a sense that one is nearing the end of life, there was a sense of urgency to make it to the other side of the learning curve:
\begin{quote}
    \sayit{I have kind of a bit of a hard charging attitude because of my age. ... I've got a couple of years to learn this stuff and after that I need to already know it. I don't want to spend the rest of my life in the learning mode of these technologies and techniques.} (L2)
\end{quote}
To save time, older adults experiencing vision loss tended to want to keep their current smartphone instead of switching. L2, a long-time Android user, talked about this in terms of optimizing learning resources, especially \sayit{from the point of view of time}: 
\begin{quote}
\sayit{Each one of those things takes a lot of effort [to learn]. Braille. Victor reader. TalkBack. All of it. O\&M and stuff. So I'm trying to minimize – preserve my learning resources in my brain by not having to learn the iPhone. I wish they would quit trying to say \say{switch to iPhone because that’s what we know.} Don't want to.} (L2)
\end{quote}
We see from the examples above how (re)learning the smartphone was not only a matter of intense labor, but it also marked a regression back into liminal space that our participants had already struggled and succeeded to break out of.

\paragraph{Resisting Conversion}
Though others quickly locked into a smartphone platform after and even before their impending vision loss, L1 explored numerous alternative smartphone models, which hindered him from learning any one of them more deeply. When he began to lose vision, he purchased an iPhone 4, since he had heard \sayit{that the iPhone is much better using Siri or the accessibility feature, and I should get an iPhone} (L1). From blind organizations, he received several phones over time: iPhone 6, iPhone SE, BlindShell Classic, and BlindShell Classic 2. L1 described using these four smartphones simultaneously, each for different purposes, rather than adopting a single model, since he didn't feel like any one phone was reliable enough, even when people around him suggested he learn one technology:
\begin{quote}
\sayit{I would prefer to have an iPhone and the BlindShell Classic 2. ... one of the women that helps me with most of the phone says I should just use the BlindShell, make phone calls on it, do everything on it, but I’m not sure if it’s good enough to do that yet. I just haven’t gotten into it enough.} (L1)
\end{quote}
The BlindShell phone is specifically designed for older BLV users, and it offers much reduced functionality as compared to the mainstream iPhone. With reduced functionality comes simplification that greatly enhances learnability and usability. Yet, adopting a phone branded for "the blind" that trades off functionality that sighted people take for granted on smart devices is entangled in the acceptance of blind identity, which L1 actively fights. As a result, L1 lacks mastery of any device and, ironically and somewhat tragically, is trapped in a state of reduced access.


\subsection{Transitioning via Mutual Aid} \label{sec_43}

Blind community support propels BLV older adults away from the ambiguity of acquiring blindness (section \ref{sec:accept_blindness}) as well as acquiring a smartphone (section \ref{sec:41acquiring_phone}). Among knowledgeable peers, they experience a sense of collective belonging and steady forward progress toward digital competency. However, some BLV older adults, especially those with later life vision loss, experience challenges gaining access to and benefiting from these resources.

\subsubsection{Going for the Phone, Gaining Community} \label{sec_431}
Participants described seeking resources to learn how to use their smartphone as a BLV person and inadvertently finding the blind community. 


Most resources–classes, webinars, ``tech talks,'' etc.–supporting BLV smartphone adoption were provided by blind advocacy groups, whose  instructors had lived experience and were therefore the best source of accurate information: \sayit{[My instructor was] very much familiar with how to teach [VoiceOver gestures] because she was a blind person} (E1). 
These resources were not only a source of basic phone know-how, but they also kept students up-to-date about the latest phone features: \sayit{[I attend] Tech Talk class where [the instructor will] look up different features that are coming out} (E4). In this way, the classes represented a step away from liminal space toward phone mastery. For L3, who was less conversant with basic phone features and jargon, these classes helped him to \sayit{talk to my daughters in the same language regarding technology} (L3). In other words, by working with the BLV community, L3 was better able to tap into his traditional social networks--i.e., family--for technical phone support.

\subsubsection{Cycle of (non)Acceptance} \label{sec_432} 
A key hindrance to BLV older adults' adoption of smartphones was the necessity of accepting one's blindness; 
without doing so, they were disconnected from the blind community and unable to tap its resources. L3 recounted when he underwent this pivotal transition:
\begin{quote}
\sayit{So here's the transition. ... I thought it was temporary. My sight's going to come back. And reality hit me. And for a while I just stopped using the damn thing [smartphone]. Because I couldn't see it. And it was just way more trouble than it was worth. ... it was the [local blind organization] that really got me going, and oh boy what classes have I taken.} (L3)
\end{quote}
Once tapped into the BLV community, smartphone learners could grow their skills, and some even became knowledgeable instructors and identity ambassadors themselves. E7, who was a teacher in the blind community, described supporting neighbors \sayit{new to losing vision} in a \sayit{low vision support group} with smartphone re-acquisition.
Beyond imparting technical skills, he instilled in them that \sayit{blindness was not a loss. We [people who were born blind] had it all the time} (E7). He endeavored to shift their perspectives on blindness \sayit{beyond the \say{I've lost my vision, I got to figure out how to get some of it back}} (E7).

We can see how engaging with the blind community around one's disability and smartphone adoption identities can kick into motion a virtuous circle that lifts one out of liminality. L4 describes this inflection point from learner to teacher as his pathway out of the \sayit{hole} of vision loss:
 \begin{quote}
 \sayit{I'll be teaching it [class] again, ... because I enjoy it. ... I'm a big believer in volunteering, and helping other people. It's just something that ... to me is one of the secrets to, or strategies for, digging your way out of a hole that you might find yourself in when you lose your vision.} (L4)
 \end{quote}
However, a vicious cycle can also occur. 
When people who acquire vision loss are unable to accept their disability identity, they are unable to tap the resources of the community and unable to fully acquire and master their smartphone. This can lead to dire consequences for their social integration and mental well-being:
\blockquote{
The first kink in the thing to make it better is getting a stream of knowledge that only comes out of the blind service delivery and the blind community systems. And if you can get that, terrific. If you can't, then you feel like you're all alone and there's nobody there to help you. And then the depression can really set in.
}{E7}

\subsubsection{The Hole of Later Life Vision Loss} \label{sec_433}
Older adults with later life vision loss felt excluded from the existing technology support system in the blind community since blind organization classes did not support their needs, introducing friction in their transition into blindness.

For some participants who were avid smartphone users before acquiring blindness, the classes were too introductory. L6 described taking an entry-level VoiceOver course that was more like \sayit{iPhone 101} (L6). The instructor demonstrated how to navigate to the settings menu, but then \sayit{proceeded to go through every category that was in the settings, and describe them and what they do} (L6). The next lecture covered \sayit{Siri ... and [had] nothing to do with VoiceOver} (L6). Suffice to say, L6 \sayit{didn't get much out of that class} (L6). 
E7 explained that the lack of classes that adequately address needs of those with later life vision loss was a matter of economics: funding allocation favors younger BLV adults who are still part of the workforce. 

There was a particular lack of support for Android users with vision loss.
Although \sayit{half of} (L2) the students in L2's classes \sayit{–especially the ones that have become non sighted later on in life–tend to have Android} (L2), he described feeling pressured \sayit{to switch to Apple} (L2) since most blind organization classes were about iPhone. 
Android classes offered by blind organizations were both scarce and not taught by Android users, which diminished their utility for older adults with vision loss. One Android instructor \sayit{confessed that ... he's only picking up this Android to teach this class} (L2).  Another instructor taught TalkBack through the lens of an iPhone:
\begin{quote}
    \sayit{He doesn't use Android. He uses iPhone. ... so he really didn't know how to use it [TalkBack]. And he said, `Well I'll learn, but let me tell you how it works on iPhone.' ... And I'm thinking, `I don't have an iPhone. I have my Android. I want to hear about how it works on Android. Talk to me about TalkBack.'} (L2)
\end{quote}
L2 described feeling unable to advocate for his needs in the blind community since he was new to the blind identity: \sayit{I sort of feel like a newcomer and like I believe in advocating for myself but I also don’t want to appear too brash. Because this is my new community, and I have to take care of it. (chuckle)} (L2). 

Due to the dominance of the iPhone in the blind community resulting in fewer support opportunities, older adult Android users in particular turned to online resources. L2 described, \sayit{I've had to look for other [online] sources other than [the local blind organization]. Like Hadley\footnote{\url{https://hadleyhelps.org/workshops/listen-with-talkback-series/android-talkback-talkback-basics}} who has the course on Androids. And different podcasts, such as the Blind Android Users Podcast\footnote{\url{https://blindandroidusers.onpodium.co/}} is very good} (L2). 


Finally, older adults with later life vision loss who were homebound experienced increased exclusion from local resources. 
For L1, in-person support is necessary \sayit{not just with the phone, but with almost everything. (chuckles)} (L1). He shared that he \sayit{didn't go through any training} (L1) for his smartphones since blind organization locations were too far away, and he felt that online resources like YouTube tutorials or manuals were too difficult, inaccessible, and incomplete to use on his own. Thus he found himself in a resource deadlock.

Without access to the blind community, older adults with later life vision loss must rely on familiar resources, such as their family. However, sighted family members may not necessarily be able to support them due to their visual understanding of smartphones:
\blockquote{If you're in a community that doesn't understand the blindness field, and you're trying to get a smartphone, you're going to try to do everything visually. And you’re going to lean on your son, your daughter, your relative, whatever. And hope that they know what’s going on. And mostly they don't because they only know how to operate it visually.}{E7}
Trying to use the smartphone visually hinders older adults with vision loss from seeking blind community support, reducing their ability to relearn their smartphone with blindness.

\section{Discussion}

\subsection{Intersecting Liminality} \label{sec_51_intersecting_liminality}

\begin{figure}[t]
    \centering
    \includegraphics[width=\linewidth]{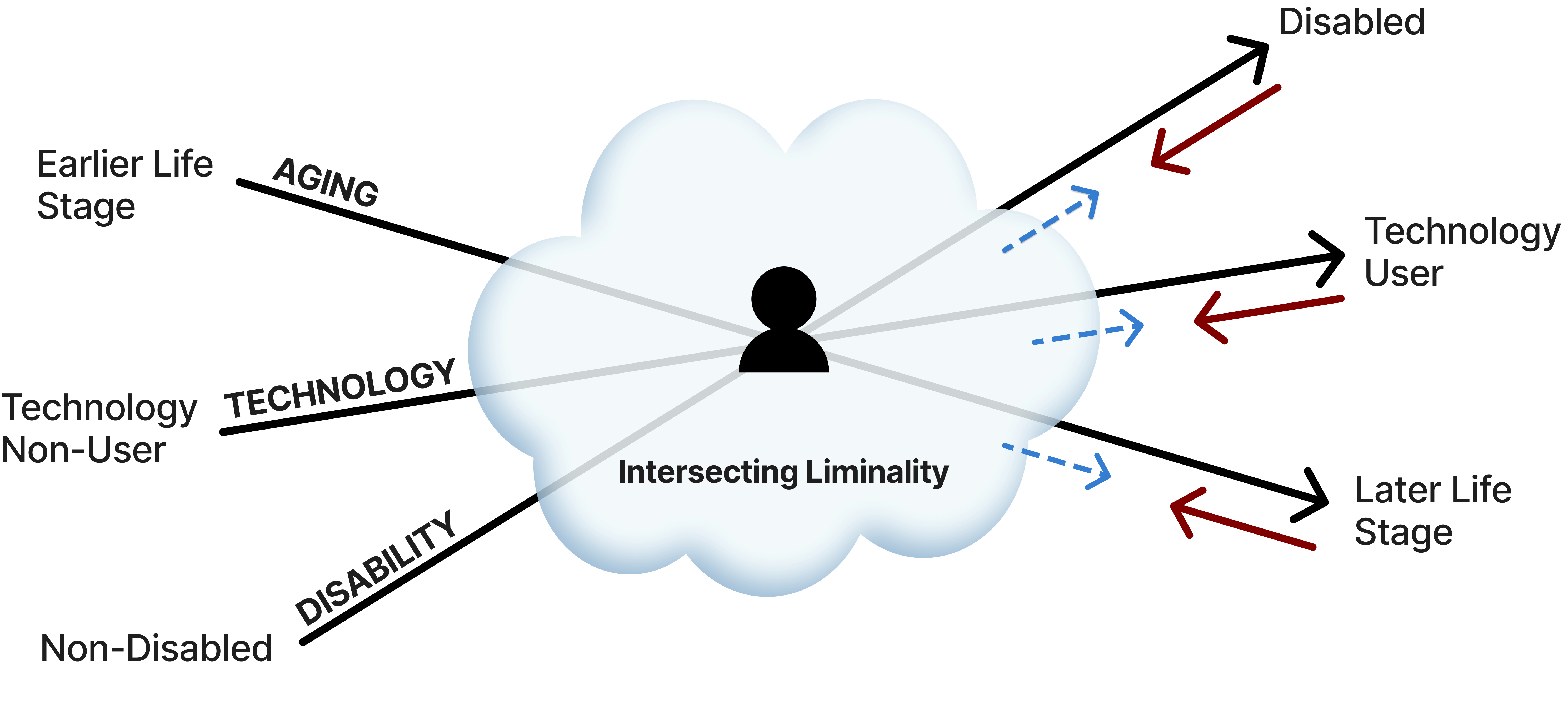}
    \caption{\textbf{Intersecting Liminality} theorizes the ways in which multiple concurrent life transitions can intersect and compound, drawing individuals back into (persistent) liminality.}
    \label{fig:intersecting_liminality}
    \Description{Intersecting Liminality. In the middle of the figure there is a cloud signifying a liminal space, and in the middle of the cloud is a person icon. Three axes go through the cloud and intersect in the middle of the cloud. Each axis represents a life transition a person might go through. First axis is aging, from an earlier life stage to a later life stage. Second axis is technology, from being a technology non-user to a technology user. The last axis is disability, the transition from being non-disabled to being disabled. Next to each axis, there is a striped blue arrow pointing out and away from the cloud and a solid red arrow pointing from out of the cloud into the liminal space cloud. The striped blue arrows point from the person and out of the cloud, signifying a person's efforts to exit the liminal space and factors that support the person to get out of the liminal space. The solid red arrows point from out of the liminal space towards the person, signifying factors that push them back into the liminal space. The person remains in the middle of the cloud, signifying that despite their efforts to exit liminality, due to the multiple intersecting life transitions, they can get stuck in liminality.}
\end{figure}

We present \textit{Intersecting Liminality}, a framework that articulates the socio-technological experiences of people who are in transition to inhabiting multiple (marginalized) identities, such as disability, older age, and technology use (refer to \autoref{fig:intersecting_liminality}). Our data reveal that for some older adults, the experiences of navigating through multiple liminalities simultaneously are fundamentally different from the experiences of navigating them individually. We argue that these liminalities intersect and their effects compound for older adults as they acquire blindness, re-acquire their smartphones, and also continue to age.

We found that participants were experiencing three concurrent life transitions of age, disability, and technology use, each with its own liminal space. 
Moreover, participants struggled to move through these spaces, as many internal and external forces pushed them back into liminality.
Prior literature documents some of the challenges and barriers to learning and mastery of smartphones and ICTs for blind adults \cite{rodrigues_getting_2015}  and blind older adults \cite{piper_technology_2017, piper_understanding_2016}.
However, our novel reading of these challenges, through the lens of multiple life transitions and liminalities, enables us to view these phenomena in new light. For example, we now understand that being blind while acquiring a smartphone can be othering and isolating because one's ambiguous identity lies outside the mainstream social practices. We also now understand how acquiring a smartphone while being blind and older can feel like being stuck in ``persistent liminality'' \cite{nicholson_living_2012} because they cannot rely on traditional social supports and have to learn amidst other challenges such as aging. 
Leveraging a framing of life transitions to understand technology acquisition can help us make sense of challenges encountered and strategies for successful adoption.



We found that challenges with acquisition may have less to do with accessibility, but more to do with resistance to accepting a BLV identity. 
Yet, prior work on acquiring technology with a disability tends to not include participant perspectives about acquiring disability at the same time as acquiring the technology \cite{piper_technology_2017, piper_understanding_2016, rodrigues_getting_2015}.
Additional to accepting a BLV identity, surprisingly, we found that those who were already smartphone users but who had newly acquired vision loss had the unique experience of having to \textit{\textbf{re-acquire}} their smartphone. 
Participants experienced a shift in identity, going from feeling tech-savvy to novice like being back in grade school (e.g., L2) due to having to relearn the smartphone as a blind person. Our participants with later life blindness already have a smartphone, but they need to learn it in a completely new way. They already have a mental model about features that are available on their smartphone, but they don't have a mental model about the language used to access it as a blind user. 
Additionally, in our field, we conceptualize adopting technology as an action a person does once, 
such as adopting an Android smartphone for the first time \cite{rodrigues_getting_2015}. When we include re-acquisition of the smartphone as a legitimate means to acquire, then we can imagine new interventions to support the relearning of technology that might occur differently and more than once due to life transitions. 
We argue that identity is consequential to acquiring technology and that we should consider re-acquisition a type of acquisition.
Thus, we recommend that researchers include people who are going through disability transitions, such as Franz et al.~\cite{franz_perception_2019}, in studies and explore how disability transitions and changes in identity interact with technology use.

The Intersecting Liminality framework stitches our findings together. Consider L2's life transitions as depicted in \autoref{fig:teaser_story}. L2 was a sighted Android smartphone user throughout his late 50s and 60s (Figure 1A). When he acquired blindness in his early 70s, he could no longer access his phone 
(Figure 1B). 
Accepting his blindness, 
he proactively sought to educate himself about how to use his Android's accessibility features, including taking classes at local blind advocacy organizations (Figure 1B, striped blue arrows pointing away from liminality cloud). Yet, he found that the blind community predominantly used iOS; even his instructor was an iOS user with minimal Android knowledge (Figure 1C, solid red arrows pointing towards liminality cloud). 
He found himself trapped in a state of constant learning, akin to that which he experienced in early life. 
L2's experience differed from those with early life blindness, who grew up “natively” within the blind community and with iOS VoiceOver. We see how L2 not only experienced liminality as a blind man who does not fit among other blind people (blindness axis), but he is also trapped in a state of persistent technology learning (smartphone axis), marking a slide back into the struggles of youth and never reaching the leisure of retirement (aging axis). To exit liminal space along the aging axis would require additionally exiting liminal space along the other two axes. When liminalities intersect, they interact and amplify the forces pushing one back into ambiguity, obscurity, and otherness, potentially ensnaring one in liminal space. 

Intersecting Liminality is akin to the concept of intersectionality \cite{crenshaw_mapping_1991}, as we begin to theorize the complex intersections and interactions between various identity-based processes. In Crenshaw's foundational work, she argues that when people–in particular black women–inhabit multiple marginalized identities, they experience gender discrimination and race discrimination in a compounding way, not separately: ``race and gender converge so that the concerns of minority women fall into the void between concerns about women's issues and concerns about racism'' (p. 1282) \cite{crenshaw_mapping_1991}. Intersecting Liminality adds dynamism to this model, attending to the transitions between identities, rather than figuring identities as static states. As such, it can account for the forward and backward progress through identity space (e.g,. moving from sighted to blind, and back again in denial) and explore the duality and vacancy of diametrically opposed identities (e.g., the state of being both novice and expert smartphone user, or neither).

Our work aligns with a growing body of intersectionality literature in HCI that calls for greater attention to nuanced and multiple marginalized identities, including class \cite{Schlesinger_intersectionalHCI_2017} and disability \cite{Harrington_Working_2023, Bennett_misrep_2021}, which are consequential for technology research and design.

\subsection{Putting Intersecting Liminality to Use}


Although we studied BLV older adults adopting smartphones, we believe that Intersecting Liminality can be used to explore multiple simultaneous life transitions. Additionally, Intersecting Liminality can help explain findings from prior work and be used for future work with people acquiring disability identities other than blindness and technologies other than smartphones.

Take for instance Franz et al.’s work with older adults with ability changes and their perceptions of and adoption of mobile device accessibility features \cite{franz_perception_2019}. Although participants with vision loss had more difficulty using their smartphones, Franz et al.~\cite{franz_perception_2019} found that before their study, many of them did not use accessibility features on their mobile devices, likely since they did not identify as disabled. The participants who did use accessibility features were disabled from birth. When the researchers followed up with participants a few weeks after the initial interview to see if they were still using accessibility features demonstrated in the first interview, many participants did not adopt accessibility features, citing reasons such as forgetting how to use features and lack of social support. That the smartphones’ accessibility features were difficult to adopt by participants experiencing age- and disability-related challenges suggests compounding effects from multiple liminalities.

As researchers and designers of assistive and accessible technologies move forward, it is important to consider that disability identity is not static; people with disabilities experience a rich range of other concurrent life transitions. These transitions are not incidental but are in fact essential to the ways in which disabled people acquire, adopt, and interact with technology. When we fail to account for the holistic, dynamic lived identities of people with disabilities at their intersections, the consequences can range from inconvenient–as when the technology prevents a BLV person from seeking support from sighted family and friends–to socially isolating–as when the technology forces them back into liminal space. 



\section{Limitations}

This study was carried out in the United States of America, thus our findings may not be applicable to other cultures or geographic areas. While 16 of our participants were blind earlier in life, only 6 had later life vision loss. Additionally, we recruited BLV older adults who use smartphones, which excludes experiences of BLV older adults who attempted to acquire a smartphone but did not continue to use one. Most (n=18) of our participants were iPhone users, while few used Android or BlindShell. This is not unexpected, since the iPhone tends to be much more prevalent among blind smartphone users \cite{griffin-shirley_survey_2017}. 
Further, since our study focuses on the experiences of BLV older adults, we are unable to make comparisons with other populations, such as BLV adults and sighted older adults. Future research should explore if Intersecting Liminality is still a productive lens in the context of, for example, texting versus voice calling preferences of these populations.

\section{Conclusion and Future Work}
In this study, we investigated how BLV older adults acquire smartphones and what resources they used to support the process of acquisition. We conducted qualitative semi-structured interviews with 22 BLV older adults, and reading through the lens of liminality, we found three main themes: (1) older adults who are blind experience liminality as they acquire smartphones, (2) later life BLV older adults experience liminality as they re-acquire a smartphone while acquiring blindness, and (3) engaging in mutual aid with the blind community supports transition between identities and out of liminality. We corroborate findings from previous literature on the challenges of technology adoption for BLV older adults. We add novel insights by analyzing our data from perspectives of life transitions and liminality (e.g., acquisition challenges stemming from resisting the acceptance of BLV identity). Finally, we introduce a framework of \textit{Intersecting Liminality}, which we recommend be applied in future work in HCI, in studies where multiple intersecting identities are involved.

\bibliographystyle{ACM-Reference-Format}
\bibliography{ASSETS2024}





\end{document}